\documentstyle[11pt,modappb,epsf,axodraw]{article} 

\newcommand{\np}[3]{{\sl Nucl. Phys.} {\bf #1} (19#2) #3}
\newcommand{\pl}[3]{{\sl Phys. Lett.} {\bf #1} (19#2) #3}
\newcommand{\pr}[3]{{\sl Phys. Rev.} {\bf #1} (19#2) #3}
\newcommand{\prl}[3]{{\sl Phys. Rev. Lett.} {\bf #1} (19#2) #3}
\newcommand{\zp}[3]{{\sl Z. Phys.} {\bf #1} (19#2) #3}
\newcommand{\ijmp}[3]{{\sl Int. J. Mod. Phys.} {\bf #1} (19#2) #3}
\newcommand{\nim}[3]{{\sl Nucl. Instrum. Meth.} {\bf #1} (19#2) #3}
\newcommand{\ibid}[3]{{\sl ibid.} {\bf #1} (19#2) #3}
\newcommand{\ej}[3]{{\bf #1} (19#2) #3}
\newcommand{\hep}[1]{{\sl hep--ph/}{#1}}

\newcommand{\WpL}{W^+_{_L}}
\newcommand{\WmL}{W^-_{_L}}
\newcommand{\ZL}{Z_{_L}}

\def\refname{\large{\bf References}}

\begin{document}
\title{PHYSICS OF THE W BOSON AT\\ 
       FUTURE LINEAR COLLIDERS\thanks{Talk and lecture presented at the 
          1997 Cracow Epiphany Conference on W boson, 4--6 January 1997,
          Krak\'ow, Poland.}
       \thanks{Research supported by a fellowship of the Royal Dutch Academy 
          of Arts and Sciences.} 
}
\author{W. Beenakker
\address{Instituut--Lorentz, University of Leiden, The Netherlands} 
}
\maketitle

\begin{abstract}
A survey is given of the various aspects of W-boson physics at the next 
generation of linear colliders. In particular, it is indicated how the W boson
can help us improve our understanding of the mechanism of mass generation and 
the structure of the non-abelian gauge-boson interactions. Also the topics of 
radiative corrections and gauge invariance are briefly addressed.  
\end{abstract}

\section{Introduction}

During the last five years the high-energy physics community has
investigated the viability of a new high-energy linear $e^+e^-$ 
collider~\cite{DESYNLC}. According to the various designs for this so-called
Next Linear Collider (NLC), an energy in the range between 500 and 1500~GeV 
seems feasible. By using the back-scattered laser-beam 
technique~\cite{backscattering}, it is possible to convert the $e^\pm$ beams 
into photon beams with comparable energy and luminosity. In this way  
the NLC could be operated as a $e^+e^-\!$, $e^-e^-\!$, $e\gamma$, or 
$\gamma\gamma$ collider. Such a versatile machine should prove an excellent 
tool in our quest to understand nature. It will provide stringent tests of the 
Standard Model (SM) of electroweak interactions~\cite{SM}, making it highly
sensitive to signals of new physics. 
If any physics beyond the SM exists, it will reveal 
itself in the production of new particles (direct signals) or in deviations in
the interactions between the SM particles (indirect signals). 

This survey is dedicated to the role played by the W boson.
In this context the most important issues to be addressed at the 
NLC are the investigation of the triple and 
quartic gauge-boson couplings, and a detailed study of the symmetry-breaking 
sector. This should shed some light on two important outstanding problems in 
present-day high-energy physics. 

The first question concerns the nature of the non-abelian interactions between
the electroweak gauge bosons. Experiments at the Tevatron and LEP1 have 
provided us with the first direct~\cite{TGCdirect} and 
indirect~\cite{TGCindirect} evidence for the existence of such interactions.
The results are, however, far from conclusive, since ${\cal O}(1)$ deviations 
from the SM couplings are still allowed. At future high-energy collider 
experiments the sensitivity to these non-abelian gauge-boson couplings will be 
increased significantly. This will either allow a verification of the SM 
couplings at the per-mil level or open a window to physics beyond the SM.

The second question that should be addressed at the NLC concerns the 
mechanism of electroweak symmetry breaking. Are the longitudinal weak 
gauge-boson modes indeed generated by means of the SM Higgs mechanism, or has
nature chosen another option? In this context there are two scenarios for the
electroweak symmetry-breaking sector. In one scenario the theory of
the fundamental interactions is assumed to be applicable to very high energies,
i.e.~up to the grand-unification scale or Planck mass 
($10^{16}$--$10^{19}$~GeV). In such a scenario (e.g.~supersymmetry) the 
symmetry-breaking sector consists of one or more elementary (pseudo-)scalar
Higgs fields, which are weakly coupled at low energies. The mass of the
lowest-lying Higgs state is predicted to be relatively small, i.e.~below 
200~GeV \cite{lowMH}. The second scenario involves the possibility of having
new strong interactions at TeV energies, related to the mechanism of mass 
generation. Such a scenario in general excludes the presence of a low-lying 
Higgs state; in fact, Higgs-like states might be absent altogether. Whatever 
the underlying theory of the strong interactions may be, these strong 
interactions will manifest themselves in the form of strongly-interacting 
longitudinal gauge bosons. After all, these longitudinal gauge-boson modes are 
a direct consequence of the mechanism of mass generation. This is reminiscent 
of the role played by the pions in hadron dynamics. In this survey the 
emphasis will be on the strongly-interacting 
scenario, since it entails the absence of direct signals at low energies. From
the viewpoint of W-boson physics this is the most interesting situation. 

In order to successfully achieve the physics goals at the NLC, a very accurate
knowledge of the SM predictions for the various observables is mandatory. It
has no use trying to perform high-precision tests of non-abelian gauge-boson 
couplings and strongly-interacting longitudinal gauge-boson interactions when 
the SM predictions do not have a matching precision. To this end a critical 
assessment is given
as to what SM ingredients are required in this respect. This involves a 
proper understanding of radiative corrections as well as a proper treatment of
finite-width effects. The weak gauge bosons are unstable particles and
experience has learned us that gauge invariance is in jeopardy when it comes to
including the finite widths of these particles. Needless to say that this can 
have large repercussions on the reliability of the SM predictions.

The outline of this survey is as follows. In section~2 a discussion is given 
of the process of longitudinal gauge-boson scattering and its intimate 
relation to the symmetry-breaking sector. In section~3 the topic of physics 
beyond the SM is addressed in a more systematic way, using the concept of 
effective Lagrangians and anomalous couplings. In section~4, finally, the SM 
predictions are considered, with emphasis on the issue of gauge invariance.
    
\section{Longitudinal gauge-boson scattering}

As mentioned before we will assume the absence of a low-lying Higgs state 
or any alternative thereof, excluding the presence of direct signals of these
states at sub-TeV energies. The most sensitive probe of the symmetry-breaking 
sector will in that case be longitudinal gauge-boson scattering, since the 
longitudinal gauge-boson modes are a direct consequence of the mechanism of 
mass generation. 

\subsection{Strong interactions between longitudinal gauge bosons in the SM}

In the SM the longitudinal gauge-boson modes are supplied by the would-be 
Nambu--Goldstone bosons, leaving behind just one elementary scalar Higgs field
(H). This is achieved by the spontaneous breakdown of the 
$SU(2)_L \times U(1)_Y$ symmetry through the Higgs mechanism. In order to 
reveal the distinctive nature of the longitudinal gauge-boson modes, high 
energies are required. At rest longitudinal ($L$) and transverse ($T$) modes 
are related by mere rotations, but at high energies they are quite different, 
since only the longitudinal modes are affected by boosts in the direction of 
flight. A simple investigation of the polarization vector $\varepsilon^\mu(k)$
of a massive gauge boson with momentum $k^\mu=(E,\vec{k}\,)$, mass $M$, and 
velocity $\beta=\sqrt{1-M^2/E^2}$ reveals: 
\begin{eqnarray}
  \varepsilon_{_T}^\mu(k) &=& (0,\vec{e}\,) \ \ \ \mbox{with}\ \ 
                              \vec{e}\cdot\vec{k}=0\ \ \mbox{and}\ \  
                              \vec{e}^{\;2}=1~, \nonumber \\
  \varepsilon_{_L}^\mu(k) &=& \frac{k^\mu}{\beta M} 
                              - \frac{M}{\beta E}\,(1,\!\vec{\,0}\,) 
                              \equiv  \frac{k^\mu}{\beta M} + V^\mu~.
\label{polvec}
\end{eqnarray}  
From this it should be clear that any amplitude involving longitudinal 
gauge bosons has the tendency to diverge at high energies as a result of 
factors $k/M \sim E/M$. In gauge theories, however, gauge cancellations take
place, resulting in properly behaved cross-sections for all energies,
i.e.~cross-sections that do not grow as a positive power of $E$. This is 
reflected by the fact that the leading-energy term $k^\mu/M$ in (\ref{polvec})
can be related by means of Ward identities to the corresponding would-be 
Goldstone mode, which is not subject to gauge cancellations.

\begin{figure}[tb]
  \begin{picture}(90,100)
    \DashLine(25,75)(75,25){4}        \Text(7,75)[lc]{$\phi^+$}
                                      \Text(88,25)[rc]{$\chi$}
    \DashLine(25,25)(75,75){4}        \Text(7,25)[lc]{$\phi^-$}
                                      \Text(88,75)[rc]{$\chi$}
  \end{picture}
  \hfill
  \begin{picture}(140,100)(0,0)
    \DashLine(50,50)(25,25){4}        \Text(7,25)[lc]{$\phi^-$}
    \DashLine(25,75)(50,50){4}        \Text(7,75)[lc]{$\phi^+$}
    \DashLine(50,50)(100,50){4}       \Text(75,57)[bc]{$H$}
    \DashLine(100,50)(125,25){4}      \Text(138,25)[rc]{$\chi$}
    \DashLine(100,50)(125,75){4}      \Text(138,75)[rc]{$\chi$}
  \end{picture}
  \hfill
  \begin{picture}(90,100)(0,0)
    \DashLine(25,75)(75,75){4}        \Text(7,75)[lc]{$\phi^+$}
                                      \Text(88,75)[rc]{$\chi$}
    \DashLine(25,25)(75,25){4}        \Text(7,25)[lc]{$\phi^-$}
                                      \Text(88,25)[rc]{$\chi$}
    \Photon(50,75)(50,25){1}{9}       \Text(35,50)[lc]{$W$}
  \end{picture}  
  \caption[]{The Feynman diagrams for the lowest-order process 
             $\phi^+\phi^- \to \chi\chi$. In the limit 
             $M_{_{W,Z}}^2 \!\ll (E^2, M_{_H}^2)$ the third diagram is 
             suppressed with respect to the other two.}
  \label{fichi}
\end{figure}
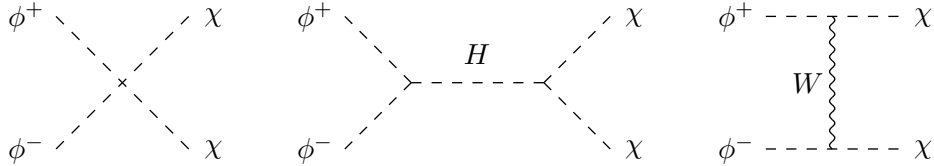

Consider now the process $\WpL\WmL \!\to \ZL\ZL$ in the limit 
$M_{_{W,Z}}^2 \!\!\ll \!(E^2\!,M_{_H}^2)$, where $E$ stands for the energy of 
the particles in
the centre-of-mass system. At this point we can make use of the equivalence 
theorem~\cite{ET}, which states that for $E^2 \gg M_{_{W,Z}}^2$ the amplitudes 
for longitudinal gauge-boson scattering are in leading-energy approximation 
equivalent to the amplitudes for the corresponding would-be Goldstone bosons. 
The terms involving the remnant $V^\mu$ occurring in (\ref{polvec}) are 
suppressed by powers of $M_{_{W,Z}}/E$.\footnote{Beyond lowest-order level 
proper care has to be taken with external self-energies and renormalization 
factors.} Hence, it suffices to study the process $\phi^+\phi^- \to \chi\chi$ 
(see Fig.\,\ref{fichi}), with $\phi^\pm$ and $\chi$ the would-be 
Goldstone bosons responsible for generating the masses of the $\mbox{W}^\pm$ 
and Z gauge bosons, respectively. 
This process is not subject to gauge cancellations and is therefore easier to 
handle. The lowest-order matrix element
is given by
\begin{equation}
  {\cal M}(\WpL\WmL \to \ZL\ZL) \approx {\cal M}(\phi^+\phi^- \to \chi\chi)
       \approx \frac{-s M_{_H}^2}{v^2(s-M_{_H}^2)}~,
\end{equation}
where $\sqrt{s}=2E$ stands for the total centre-of-mass energy and $v=246$~GeV
for the electroweak symmetry-breaking scale.

If the Higgs boson is very heavy ($E^2\ll M_{_H}^2$)
an interesting phenomenon occurs. Naively the amplitude for the process
$\WpL\WmL \to \ZL\ZL$ is expected to diverge like $E^4$, but after the gauge 
cancellations have taken place an $E^2$ behaviour remains. To be more precise: 
${\cal M} \to s/v^2$. This behaviour is purely a consequence of the fact that 
the symmetry is broken and does not contain any information on the dynamics 
responsible for the symmetry breaking, i.e.~the Higgs. It is 
prescribed by the Low-Energy Theorem (LET)~\cite{LET} corresponding to the 
broken symmetry, which states that the would-be Goldstone bosons decouple at 
low energies (up to gauge and Yukawa couplings). We will come back to this 
point later on. Obviously something must happen in the TeV range in order to 
salvage unitarity, after all the SM is unitary by construction; in other words,
strong-interaction effects are expected in that regime. Indeed, in the 
heavy-Higgs approximation the quartic Higgs self couplings 
$\lambda \propto M_{_H}^2/v^2$ are large and higher-order corrections
$\propto M_{_H}^2/(4\pi v)^2$ can not be discarded. Two natural scales govern
the dynamics of these strong-interaction effects: $M_{_H}$ (resonance) and 
$4\pi v \sim 3$~TeV (corrections). The Goldstone bosons, which decouple at 
low energies, will start to interact strongly if $E={\cal O}(M_{_H},4\pi v)$. 
As a result of the equivalence theorem, the same holds for the longitudinal 
gauge bosons. 

Recapitulating: the LET behaviour of longitudinal gauge-boson scattering at 
intermediate energies ($M_{_{W,Z}}^2\ll E^2\ll M_{_H}^2$) is a signature of a 
strongly-interacting symmetry-breaking sector, since it would be absent if the
Higgs boson were to be light. In contrast, the longitudinal gauge-boson 
interactions do not yet appear to be strong at these energies. The actual 
strong dynamics only shows up when the energy approaches the realm of the
symmetry-breaking sector.

\subsection{A systematic analysis of longitudinal gauge-boson scattering}

We can now turn to a more general discussion of a strongly-interacting 
symmetry-breaking sector. To this end all other interactions (like weak and 
Yukawa interactions) are for the moment simply neglected. Two guiding 
principles are relevant for the discussion. 

The first thing to note is that the SM Higgs sector has a larger symmetry than 
just the local $SU(2)_L \times U(1)_Y$. The Higgs Lagrangian
\begin{equation}
  {\cal L}_{_H} = \left(\partial_\mu \Phi\right)^\dagger
       \left(\partial^\mu \Phi\right)
       - \lambda \left(\Phi^\dagger \Phi -\frac{\mu^2}{2\lambda}\right)^2~,
\end{equation}
with $\phi$ the complex Higgs doublet, is also symmetric under global
$SU(2)_L \times SU(2)_R$ transformations. This corresponds to the symmetry 
under rotation of the four components. After the spontaneous symmetry breaking
has taken place only the symmetry under rotation of the would-be Goldstone 
bosons is apparent. This (isospin) symmetry is usually referred to as the 
custodial $SU(2)_V$ symmetry. As far as strong interactions are concerned the 
same symmetry applies to the weak gauge bosons, being related to the would-be 
Goldstone bosons by the equivalence theorem. In this context both
$(W^+\!,Z,W^-)$ and $(\phi^+\!,\chi,\phi^-)$ behave as isospin triplets 
($I=1$). In the SM this symmetry is (weakly) broken by hypercharge 
interactions. This leads to the relation 
$M_{_W} = M_{_Z}\,\cos\theta_w \neq M_{_Z}$, with $\theta_w$ the weak mixing 
angle. The fact that the $SU(2)_V$ symmetry is not broken by the strong
symmetry-breaking interactions is reflected by the observation that the
so-called $\rho$ parameter, representing the relative 
strength of neutral- and charged-current interactions at low energies,
is close to unity: $\rho = 1 + {\cal O}(\%)$.
In view of the strong experimental restrictions on this $\rho$ parameter, any 
symmetry-breaking mechanism other than the one adopted in the SM should better 
obey the custodial $SU(2)_V$ symmetry.

The second guiding principle is provided by the LET corresponding to the 
spontaneous symmetry-breaking mechanism. The breaking of the (axial)
symmetry, being of the order of the masses of the gauge bosons, is only a weak
one when compared with
the scale governing the strongly-interacting sector. In order to assess the
implications of this observation, we first redefine the Higgs doublet by 
representing the would-be Goldstone bosons by a non-linear realization of the 
full $SU(2)_L \times SU(2)_R$ symmetry group:
\begin{equation}
 \Phi = \left(\begin{array}{c} \phi^+ \\ \frac{v+H+i\chi}{\sqrt{2}} 
              \end{array}\right) = \Sigma\left(\begin{array}{c} 0 \\ 
                                   \frac{v+H'}{\sqrt{2}} \end{array}\right)~.
\end{equation}
Here $\Sigma=\exp(i\omega^j\tau^j/v)$ transforms as 
$\Sigma \to U_L \Sigma U_R^\dagger$ under $SU(2)_L \times SU(2)_R$, with
$U_{L,R} \in SU(2)$ and $\tau^j$ \,($j=1,2,3$) the Pauli matrices. For energies
well below $M_{_H}$ the heavy Higgs field can be integrated out, resulting in
a non-renormalizable chiral Lagrangian:
\begin{equation}
  {\cal L}_{_H} = \frac{v^2}{4}\,\mbox{Tr}\left([\partial_\mu \Sigma^\dagger]
                  [\partial^\mu \Sigma]\right)
                  + \mbox{two terms with four derivatives} + \cdots
  \label{chiral}
\end{equation}
In this limit the would-be Goldstone fields $\omega^j$ are related to the 
original $\phi^\pm$ and $\chi$ fields according to 
\begin{eqnarray}
  \omega^j &=& \phi^j \:\sum_{k=0}^\infty \frac{(2k)!}{2^{2k} (k!)^2 (2k+1)}
               \left( \frac{\phi^l \phi^l}{v^2} \right)^k~, \nonumber \\
  \phi^\pm &=& \frac{\phi^1 \mp i\phi^2}{\sqrt{2}}\ \ ,\ \ \chi = \phi^3~.
\end{eqnarray}
The chiral Lagrangian (\ref{chiral}) represents the effective interactions
between the would-be Goldstone bosons in the heavy-Higgs limit. It only 
involves derivative couplings, since $\Sigma^\dagger \Sigma =1$.
As a consequence there 
will be no strong scattering at low energies, i.e.~the would-be Goldstone 
bosons decouple at low energies (up to gauge and Yukawa couplings).
The first (kinetic) term in (\ref{chiral}), with the lowest number of 
derivatives, is universal. Its coefficient is fixed by the electroweak 
symmetry-breaking scale $v$. The terms
in the expansion with a larger number of derivatives are linked to the 
dynamics of the symmetry-breaking sector. They are suppressed by factors
$E^2/(4\pi v)^2$, with $4\pi v$ the generic scale for the strong
interactions (and resonances). In the case of the SM these terms will contain 
information on the Higgs sector. In general strongly-interacting
scenarios the above chiral Lagrangian parametrizes the dynamics under the
assumption of $SU(2)_L \times SU(2)_R$ symmetry, with the $\omega^j$ the 
would-be Goldstone bosons responsible for the generation of the gauge-boson
masses. 

What can we learn from longitudinal gauge-boson scattering, bearing in mind the
above guiding principles for a general strongly-interacting symme\-try-breaking
sector? Applying the equivalence theorem, this is equivalent to an analysis of 
the generic process 
\begin{equation}
   \phi^i(p_i) + \phi^j(p_j) \to \phi^k(p_k) + \phi^l(p_l)~,
\end{equation} 
involving identical massless spinless particles (as far as the strong 
interactions are concerned). Exploiting crossing symmetry 
and Bose symmetry for identical particles, the corresponding matrix element 
can be written as
\begin{equation}
  {\cal M} = A(s,t,u)\,\delta_{ij}\delta_{kl} 
           + A(t,s,u)\,\delta_{ik}\delta_{jl} 
           + A(u,t,s)\,\delta_{il}\delta_{jk}~, 
\end{equation}
with $A(s,t,u)=A(s,u,t)$. Here we introduced the standard Mandelstam variables 
$s=(p_i+p_j)^2$, $t=(p_i-p_k)^2=-s(1-\cos\theta)/2$, 
and $u=(p_i-p_l)^2=-s(1+\cos\theta)/2$, with $s+t+u=0$ and
$\theta = \angle(\vec{p}_i,\vec{p}_k)$. Projection on the elastic (s-channel) 
isospin eigenstates yields:
\begin{eqnarray}
  I=0:& & {\cal M}_0 = 3A(s,t,u)+A(t,s,u)+A(u,t,s)~, \nonumber \\
  I=1:& & {\cal M}_1 = A(t,s,u)-A(u,t,s)~, \nonumber \\
  I=2:& & {\cal M}_2 = A(t,s,u)+A(u,t,s)~.
  \label{MI}
\end{eqnarray} 
This can be rewritten in terms of charge eigenchannels:
\begin{eqnarray}
  {\cal M}(\phi^+\phi^- \to \phi^+\phi^-) &=& A(s,t,u) + A(t,s,u) = 
          \frac{1}{3}{\cal M}_0 + \frac{1}{2}{\cal M}_1 
        + \frac{1}{6}{\cal M}_2~, \nonumber \\
  {\cal M}(\phi^+\phi^- \to \chi\chi)\hphantom{Aa} &=& A(s,t,u) = 
          \frac{1}{3}{\cal M}_0 - \frac{1}{3}{\cal M}_2~, \nonumber \\
  {\cal M}(\chi\,\chi \to \chi\,\chi)\hphantom{AAA} &=& A(s,t,u) + A(t,s,u) 
                                                       + A(u,t,s) =
          \frac{1}{3}{\cal M}_0 + \frac{2}{3}{\cal M}_2~, \nonumber \\
  {\cal M}(\phi^\pm\chi \to \phi^\pm\chi)\hphantom{Aa} &=& A(t,s,u) = 
          \frac{1}{2}{\cal M}_1 + \frac{1}{2}{\cal M}_2~, \nonumber \\
  {\cal M}(\phi^\pm\phi^\pm \to \phi^\pm\phi^\pm) &=& A(t,s,u) + A(u,t,s) = 
          {\cal M}_2~. 
\end{eqnarray}

For energies well below the scale of the strong symmetry-breaking interactions
the LET predicts $A(s,t,u)=s/v^2 + {\cal O}(s^2/[16\pi^2 v^4])$. So, the 
amplitudes vanish at low energies (up to gauge and Yukawa couplings) with 
fixed slope at $s=0$. The ${\cal O}(s^2/[16\pi^2 v^4])$ terms contain 
information on the symmetry-breaking dynamics.

The amplitudes for the elastic (s-channel) isospin eigenstates can 
be projected on partial waves:
\begin{equation}
  {\cal M}_I = 32\pi\sum_{J=0}^{\infty}(2J+1)\,a_{_{IJ}}(s)\,
               P_{_J}(\cos\theta)~,
\end{equation}
with $P_{_J}(\cos\theta)$ the Legendre polynomials and $J$ the total angular 
momentum. Elastic unitarity reads $\mbox{Im}(a_{_{IJ}}) = |a_{_{IJ}}|^2$ or 
$\mbox{Im}(a_{_{IJ}}^{-1}) = -1$ for all individual channels. 
This can be solved in terms of phase shifts: 
$a_{_{IJ}} = \sin\delta_{_{IJ}} \exp(i\delta_{_{IJ}})$, spanning the unitarity
circle. In the presence of inelastic channels, like 
$\phi\phi \to 4\phi$, the $a_{_{IJ}}$ are required to lie inside
the unitarity circle, i.e.~$\mbox{Im}(a_{_{IJ}}) > |a_{_{IJ}}|^2$. These
inelastic channels are suppressed in the energy expansion, since they only 
contribute at ${\cal O}(s^4/[4\pi v]^8)$. They are only relevant for energies 
above 2~TeV and are therefore neglected in the following.
The LET predicts the lowest-order partial waves to be fixed:
\begin{equation}
  a_{_{00}} = \frac{s}{16\pi v^2}\ \ ,\ \ a_{_{11}} = \frac{s}{96\pi v^2} 
  \ \ ,\ \ a_{_{20}} = -\frac{s}{32\pi v^2}~.
\end{equation}
Note that because of Bose symmetry $I+J$ should be even. The isoscalar 
($I=J=0$) and isovector ($I=J=1$) partial waves are attractive, leaving open 
the possibility of finding a resonance at high energies in those channels. 
The partial wave for $I=2$ and $J=0$ is repulsive, excluding the presence of
resonances.

In higher order in the energy expansion the partial waves are not universal 
anymore and two unknown parameters show up~\cite{chiralho}: 
($\varepsilon \downarrow 0$)
\begin{eqnarray}
  A(s,t,u) &=& \frac{s}{v^2} + \frac{1}{16\pi^2 v^4}\left[ \beta_1(\mu^2)\,s^2 
               + \beta_2(\mu^2)\,tu  
               - \frac{1}{2}s^2\log\left(\frac{-s-i\varepsilon}{\mu^2}\right) 
               \right. \nonumber \\
           & & \left. {} 
               -\frac{1}{6}t(t-u)\log\left(\frac{-t-i\varepsilon}{\mu^2}\right)
               -\frac{1}{6}u(u-t)\log\left(\frac{-u-i\varepsilon}{\mu^2}\right)
               \right]~.\hphantom{AA}
  \label{Astu}
\end{eqnarray}
Here the arbitrary parameter $\mu$ is merely introduced to make the arguments
of the logarithms dimensionless.
These logarithmic terms are a direct consequence of analyticity and elastic 
partial-wave unitarity for the $a_{_{IJ}}$. In the language of chiral 
Lagrangians these logarithms are the result of chiral one-loop effects.
For the renormalization of the one-loop effects the two terms in the chiral 
Lagrangian with four derivatives are required. This explains the occurrence of
the unknown coefficients $\beta_{1,2}$, related to the dynamics of the model. 
In the same way the presence of the Higgs is required in the
SM for having a renormalizable theory. The ${\cal O}(s^2/[16\pi^2 v^4])$ 
corrections to the partial-wave amplitudes $a_{_{IJ}}$ can be obtained from
(\ref{MI}) and (\ref{Astu}) by an appropriate projection of the matrix 
elements ${\cal M}_I$.

It turns out that the coefficients $\beta_{1,2}$ contribute with different 
signs for isoscalar and isovector resonances~\cite{Hikasa}. An isoscalar 
resonance gives rise to a positive contribution to $a_{_{00}}$ and a negative 
one to $a_{_{11}}$. The reverse happens for an isovector resonance. As a 
result, the change in the slope of the partial waves provides crucial 
information on the strong dynamics, allowing a disentangling of the different
 models.

\subsection{Experimental sensitivity at the NLC}

\begin{figure}[tb]
  \begin{picture}(170,120)(5,-25)
    \ArrowLine(35,80)(60,80)        \Text(20,80)[lc]{$e^-$}
    \ArrowLine(60,20)(35,20)        \Text(20,20)[lc]{$e^+$}
    \ArrowLine(60,80)(95,80)        \Text(83,87)[bc]{$e^-/\nu_e$}
    \ArrowLine(95,20)(60,20)        \Text(83,15)[tc]{$e^+/\bar{\nu}_e$}
    \Photon(60,80)(95,50){1}{7}     \Text(73,65)[tr]{$L$}
    \Photon(60,20)(95,50){1}{7}     \Text(73,36)[br]{$L$}
    \Photon(95,50)(125,75){1}{6}    \Text(117,65)[tl]{$L$}
    \Photon(95,50)(125,25){1}{6}    \Text(117,36)[bl]{$L$}
    \ArrowLine(150,60)(125,75)      \Text(165,60)[rc]{$\bar{f}_1'$}
    \ArrowLine(125,75)(150,90)      \Text(165,90)[rc]{$f_1$}
    \ArrowLine(150,10)(125,25)      \Text(165,10)[rc]{$\bar{f}_2'$}
    \ArrowLine(125,25)(150,40)      \Text(165,40)[rc]{$f_2$}
    \GCirc(95,50){10}{1}
    \Text(80,-10)[]{(a)}
  \end{picture} 
  \hfill 
  \begin{picture}(175,120)(0,-25)
    \ArrowLine(60,50)(35,25)        \Text(20,25)[lc]{$e^+$}
    \ArrowLine(35,75)(60,50)        \Text(20,75)[lc]{$e^-$}
    \PhotonArc(85,50)(25,0,360){1}{20}
    \Text(85,83)[bc]{$W_L^-$}
    \Text(85,22)[tc]{$W_L^+$}
    \PhotonArc(135,50)(25,90,270){1}{10}
    \Text(132,83)[br]{$W_L^-$}
    \Text(132,22)[tr]{$W_L^+$}
    \ArrowLine(160,10)(135,25)      \Text(175,10)[rc]{$\bar{f}_2'$}
    \ArrowLine(135,25)(160,40)      \Text(175,40)[rc]{$f_2$}
    \ArrowLine(160,60)(135,75)      \Text(175,60)[rc]{$\bar{f}_1'$}
    \ArrowLine(135,75)(160,90)      \Text(175,90)[rc]{$f_1$}
    \GCirc(60,50){8}{0.5}
    \GCirc(110,50){10}{1}
    \Text(85,-10)[]{(b)}
  \end{picture}
  \caption[]{The (a) fusion and (b) $W$-rescattering mechanisms for 
             longitudinal gauge-boson scattering in $e^+e^-$ collisions. 
             The open circle connecting the longitudinal gauge-boson 
             propagators represents the strong interactions.}
  \label{LLtoLL}
\end{figure}
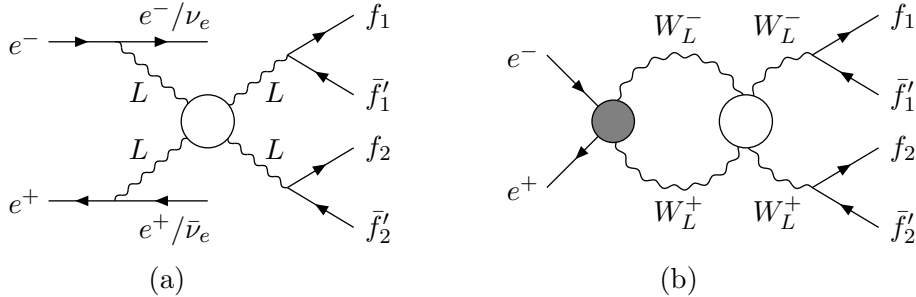

At the NLC the longitudinal gauge-boson scattering processes show up in two
distinct ways. The first one, displayed in Fig.\,\ref{LLtoLL}a, involves the 
emission of massive gauge bosons (mainly W bosons) from the initial-state 
electron and positron. 
These gauge bosons subsequently interact with each other. This mechanism is 
called gauge-boson fusion. The advantage of this mechanism is the possibility 
to access all different channels (in isospin, angular momentum, and charge), 
especially if also the $e^-e^-$ mode of the collider is used. By analyzing the
invariant-mass distributions in the various channels, one should be able to 
differentiate between the various mass-generation models, provided the 
experimental resolution in the hadronic channels is sufficient for 
distinguishing between hadronic decays of W and Z bosons. The possibility to 
polarize the initial-state beams can be exploited to enhance the sensitivity, 
by increasing the number of W bosons emitted from the initial-state $e^\pm$. 
The main drawback of the fusion mechanism is
the inefficient use of the collider energy, owing to the spectator leptons that
carry away part of the energy. This explains why one has to resort to
the invariant-mass distributions of the produced gauge bosons in order to 
investigate the dynamics of the symmetry-breaking sector; after all,
the energy transmitted to the longitudinal gauge bosons is not fixed.
As a result of the inefficient use of the collider energy, the signal 
cross-sections are relatively small until the actual resonances are formed.

The second and most promising way of studying longitudinal gauge-boson 
scattering at the NLC involves a detailed investigation of the process 
$e^+e^- \to W^+W^-$. The produced W bosons can trigger strong final-state 
interactions, called rescattering (see Fig.\,\ref{LLtoLL}b).
In view of angular-momentum conservation $J=0$ is not allowed and only the 
$I=J=1$ channel is accessible in this reaction. Making use of the elastic
unitarity conditions for the strong final-state interactions
in this isovector channel, the rescattering can be represented by a simple 
Mushkelishvili--Omn\`es form factor: ($\varepsilon \downarrow 0$)
\begin{equation}
  {\cal M}(e^+e^- \!\to W^+W^-) = {\cal M}^{(0)}
          \exp\left[\frac{s}{\pi}\int_0^\infty ds'\,
                    \frac{\delta_{_{11}}(s')}{s'(s'-s-i\varepsilon)}\right]~.
\end{equation}
Here $\delta_{_{11}}(s)$ stands for the isovector phase shift and 
${\cal M}^{(0)}$ indicates the lowest-order W-pair 
production amplitude without final-state interactions. When energy loss 
through initial-state photon radiation is kept under control, this rescattering
process involves a relatively well-defined energy. The sensitivity to
the interesting longitudinal gauge-boson modes can be enhanced by cutting away
W bosons that are produced in the forward direction, which are predominantly 
transversely polarized. In addition, the angular distributions of the decay 
products of the W bosons can be exploited to increase the sensitivity to 
the longitudinal polarization states\footnote{It should be noted 
that the fermionic currents associated with these decays have (roughly) the 
same properties as the polarization vectors in (\ref{polvec}), since a large 
majority of the decaying time-like gauge bosons is close to being on-shell.}.
In order to have access to the 
Higgs-like $I=J=0$ isoscalar channel at the NLC, one has to resort to the 
$\gamma\gamma$ collider mode with polarized photon beams. In this mode the 
isoscalar interactions can be investigated in the rescattering process 
$\gamma\gamma \to W^+W^-$ for energies comparable to the ones attainable with 
the $e^+e^-$ mode. The overwhelming production of transverse W bosons, however,
completely swamps the interesting rescattering phenomena. In this respect the 
process $\gamma\gamma \to ZZ$ looks more promising, but even there the 
transversely polarized Z bosons seriously hamper the study of a 
strongly-interacting isoscalar sector.

A recent study of the above processes~\cite{Barklowstrong}, taking into 
account effects from initial-state radiation and beamstrahlung, has shown that 
a 500~GeV NLC with an integrated luminosity of 80\,$\mbox{fb}^{-1}$ will allow 
to exclude isovector resonances up to 2.5~TeV or discover such a resonance 
up to 1.5~TeV. A 1.5~TeV NLC with an integrated luminosity of 
190\,$\mbox{fb}^{-1}$ should be able to compete with the Large Hadron Collider
(LHC) in the isoscalar and non-resonant channels. In the isovector channel
conclusive statements are expected: a strongly-interacting 
symmetry-breaking sector will be clearly distinguishable from the SM with a 
light Higgs, even if the associated isovector resonance has a very large mass.
Even more, it will be possible to make statements concerning the mass of the 
isovector resonance. For instance, a 4~TeV resonance is expected to be 
distinguishable from an infinitely heavy `resonance'. Based on these
assessments it is safe to state that the NLC will be a prime machine for 
probing the symmetry-breaking sector, in particular if nature has chosen a 
strongly-interacting isovector scenario. 
 
\section{Physics beyond the Standard Model}

The previous section has been exclusively dedicated to the mechanism of mass
generation and its measurable effects at the NLC through longitudinal 
gauge-boson scattering. One may, however, ask oneself the question how any
new physics (NP) beyond the SM will manifest itself. The most obvious signal
would be the direct production of the particles associated with this NP 
sector. For this to happen the collider energy should be above the threshold 
for the production of these particles. If this is not the case the NP sector 
can only reveal itself indirectly, i.e.~through deviations in the interactions 
between `established' (SM) particles. These deviations are generally referred 
to as anomalous interactions.
A parametrization of these anomalous interactions can be achieved by
introducing the concept of effective Lagrangians. 

\subsection{The concept of effective Lagrangians and anomalous couplings}

Let us assume that the energy
scale associated with the NP sector (e.g.~particle masses) is given by 
$\Lambda_{NP}$ and that this energy scale largely exceeds the available
collider energy. Then the NP effects will manifest themselves in 
two distinct (indirect) ways. \vspace*{2mm}

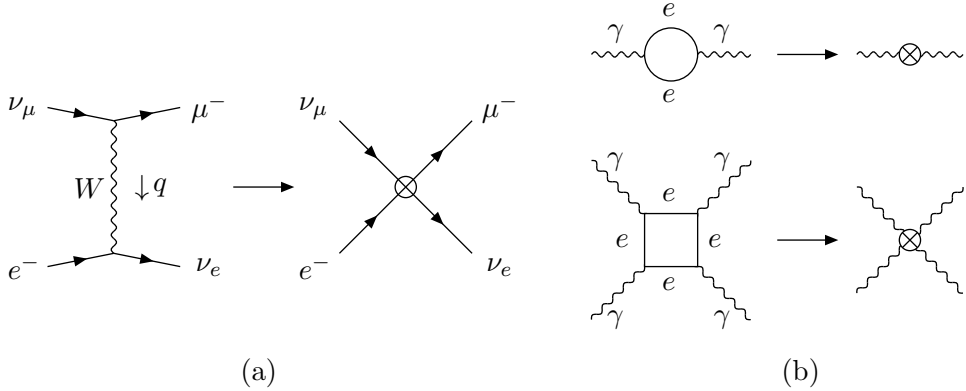
\begin{figure}[tb]
  \begin{picture}(355,150)(0,-25)
    \ArrowLine(15,90)(40,85)          \Text(0,90)[lc]{$\nu_\mu$}
    \ArrowLine(40,85)(65,90)          \Text(83,90)[rc]{$\mu^-$}
    \Photon(40,85)(40,35){1}{9}       \Text(25,60)[lc]{$W$}
                                      \Text(48,60)[lc]{$\downarrow$}
                                      \Text(55,60)[lc]{$q$}
    \ArrowLine(15,30)(40,35)          \Text(0,30)[lc]{$e^-$}
    \ArrowLine(40,35)(65,30)          \Text(81,30)[rc]{$\nu_e$}
    \Line(85,60)(105,60)             
    \ArrowLine(105,60)(106,60)
    \GCirc(150,60){4}{1} 
    \ArrowLine(125,35)(150,60)        \Text(110,30)[lc]{$e^-$}
    \ArrowLine(150,60)(175,35)        \Text(191,30)[rc]{$\nu_e$}
    \ArrowLine(125,85)(150,60)        \Text(110,90)[lc]{$\nu_\mu$}
    \ArrowLine(150,60)(175,85)        \Text(193,90)[rc]{$\mu^-$}
    \Text(95,-10)[]{(a)}
    \Photon(220,110)(280,110){1}{12}  \Text(230,117)[bc]{$\gamma$} 
                                      \Text(270,117)[bc]{$\gamma$}  
    \GCirc(250,110){10}{1}            \Text(250,125)[bc]{$e$}
                                      \Text(250,97)[tc]{$e$}
    \Line(290,110)(310,110)
    \ArrowLine(310,110)(311,110)
    \Photon(320,110)(360,110){1}{8}
    \GCirc(340,110){4}{1}              
    \Line(337.2,112.8)(342.8,107.2)
    \Line(337.2,107.2)(342.8,112.8)
    \Photon(220,70)(280,10){1}{15}    \Text(233,70)[rc]{$\gamma$}
                                      \Text(267,70)[lc]{$\gamma$}
    \Photon(220,10)(280,70){1}{15}    \Text(233,10)[rc]{$\gamma$}
                                      \Text(267,10)[lc]{$\gamma$}
    \BBoxc(250,40)(20,20)             \Text(250,55)[bc]{$e$}
                                      \Text(250,26)[tc]{$e$}
                                      \Text(235,40)[rc]{$e$}
                                      \Text(265,40)[lc]{$e$}
    \Line(290,40)(310,40)
    \ArrowLine(310,40)(311,40)
    \Photon(320,60)(360,20){1}{10}
    \Photon(320,20)(360,60){1}{10}
    \GCirc(340,40){4}{1}              
    \Line(337.2,42.8)(342.8,37.2)
    \Line(337.2,37.2)(342.8,42.8)
    \Text(300,-10)[]{(b)}     
  \end{picture}  
  \caption[]{Examples of (a) the exchange of heavy NP particles and 
             (b) integrating out heavy NP particles in loops.}
  \label{NPeffects}
\end{figure}

{\it -- Exchange of heavy NP particles:}
An example of an effective Lagrangian of this type is provided by the (pre-SM)
Fermi contact interactions, describing the V$-$A structure of weak 
interactions at low energies. For example, the effective Lagrangian for the 
charged-current interactions between electrons, muons, and their neutrinos 
reads
\begin{equation}
  {\cal L}_{\mbox{\tiny CC}}^{e,\mu} = -\frac{G_F}{\sqrt{2}}\,
                                       J^\dagger_\lambda J^\lambda~,
\end{equation}
with
\begin{equation} 
  J^\lambda = \overline{\Psi}_{\nu_e} \gamma^\lambda (1-\gamma_5) \Psi_e
            + \overline{\Psi}_{\nu_\mu} \gamma^\lambda (1-\gamma_5) \Psi_\mu~.
\end{equation}
This effective Lagrangian is based on $U(1)$-invariant fermionic currents, as
motivated by the (at that time) established theory of electromagnetic 
interactions. The Lagrangian ${\cal L}_{\mbox{\tiny CC}}^{e,\mu}$ is 
evidently not renormalizable.
It is of dimension six, or in other words, the coupling constant $G_F$ has 
dimension $(\mbox{mass})^{-2}$. The cross-sections of reactions described by 
this effective Lagrangian seem to violate unitarity at high energies,
e.g.~$\sigma(\nu_\mu e^- \to \mu^- \nu_e) \sim G_F^2 s$. At high energies the 
underlying theory, i.e.~the SM, will come to the rescue. To be more precise,
the reaction is caused by the exchange of a spin-1 particle, 
called W boson (see Fig.\,\ref{NPeffects}a). 
The Fermi constant $G_F$ will turn into a form factor 
$\sqrt{2}\,e^2/[8\sin^2\theta_w(M_{_W}^2-q^2)]$, ensuring that 
the cross-sections have a proper high-energy behaviour. From the low-energy 
limit $|q^2|\ll M_{_W}^2$ one can deduce 
$G_F = \sqrt{2}\,e^2/[8\sin^2\theta_w M_{_W}^2]$. 
At these energies the charged-current 
interactions appear weak, whereas for $|q^2| = {\cal O}(M_{_W}^2)$ the
full dynamics related to the SM W boson shows up and the electromagnetic and
weak interactions become of comparable strength. Note also that in the above 
effective Lagrangian the concept of parity conservation, valid for the 
electromagnetic interactions, has been abandoned. \vspace*{2mm}

{ \it -- Integrating out heavy NP particles appearing in loops:}
As an example we could go back to the pre-QED time. In this setting one could 
consider anomalous 
interactions between photons, caused by the interaction between these photons 
and unknown NP particles (called electrons). For energies $E_\gamma \ll m_e$ 
these low-energy interactions can be cast into an effective Lagrangian based 
on the free-photon field strength 
$F_{\mu\nu} = \partial_\mu A_\nu-\partial_\nu A_\mu$:
\begin{eqnarray}
  {\cal L}_{\mbox{\scriptsize eff}} &=& -\frac{1}{4}F^{\mu\nu}F_{\mu\nu}
      + {\cal L}_{\mbox{\scriptsize gauge fixing}}
      + \frac{\beta_1 e^2}{16\pi^2 m_e^2} \left[ F^{\mu\nu}\Box F_{\mu\nu}
      + \frac{\epsilon_1}{m_e^2} F^{\mu\nu}\Box^2 F_{\mu\nu} \right] 
      \nonumber \\
                           & &
      {}+ \frac{e^4}{16\pi^2 m_e^4}\left[ \beta_2 (F^{\mu\nu}F_{\mu\nu})^2
      + \beta_3 (F_{\mu\nu}\varepsilon^{\mu\nu\rho\lambda}F_{\rho\lambda})^2
      \right] + \cdots
\end{eqnarray}
Again we end up with a Lagrangian that is non-renormalizable in finite order,
i.e.~at each order in the energy expansion the ultraviolet divergences can
only be cancelled by introducing additional terms of higher order. The 
coefficients $\beta_i$ and $\epsilon_i$ parametrize the deviations from the 
standard interactions and are ordered according to the expansion in 
$E_\gamma/m_e$ or, equivalently, according to the dimension of the terms in 
the effective Lagrangian. At low energies the various terms in the effective 
Lagrangian seem to jeopardize unitarity. At higher energies, however, the 
dynamics of QED will show up, turning the coefficients into form factors and 
giving distinct predictions for the low-energy limits. These form factors will
guarantee a proper high-energy behaviour and will be {\it related} according 
to the underlying theory (ensuring the renormalizability).
Note that a factor $1/(16\pi^2)$ appears each time a heavy particle is 
integrated out in a loop. For instance, $\beta_1$ corresponds to the
electron-loop contribution to the vacuum polarization, and $\beta_{2,3}$ 
correspond to the electron-loop contribution to light-by-light scattering
(see Fig.\,\ref{NPeffects}b). The coefficient $\epsilon_1$ corresponds to the 
energy expansion of the heavy-particle propagators appearing in the vacuum 
polarization and has accordingly no extra factor $1/(16\pi^2)$.\vspace*{2mm}

Bearing in mind the above examples, it is possible to construct a 
non-renormalizable effective Lagrangian describing physics beyond the SM. 
This effective Lagrangian takes the general form
\begin{equation}
  {\cal L}_{\mbox{\scriptsize eff}} = {\cal L}_{\mbox{\tiny SM}} 
                                      + {\cal L}_{\mbox{\tiny NR}}
  \ \ ,\ \ {\cal L}_{\mbox{\tiny NR}}  = \sum_{n=5}^{\infty} \sum_i 
  \frac{\alpha_i^{(n)}}{\Lambda_{NP}^{n-4}} O_i^{(n)}~,
\end{equation}
with $n$ the dimension of the interaction, $\alpha_i^{(n)}$ the dimensionless
anomalous couplings, and $O_i^{(n)}$ the operators describing the anomalous
interactions
between the `established' particles. These operators respect the symmetry of 
the SM, i.e.~they are invariant under the local $SU(2)_L\times U(1)_Y$ gauge 
transformations and depend only on covariant derivatives and field strengths. 
It should be noted that certain global symmetries that are present in the SM
need not be maintained in the effective Lagrangian.

There are two scenarios for such an effective Lagrangian. In the first 
(linear) scenario the SM is completely established, including the presence of 
a light scalar Higgs boson. In that case the decoupling 
theorem~\cite{decoupling} applies. This theorem states that if the heavy NP 
particles do not acquire their masses by means of the SM Higgs mechanism, then 
$\alpha_i^{(n)}$ does not depend on $\Lambda_{NP}$. Consequently the 
dimension-$n$ operators are suppressed by factors $(Q/\Lambda_{NP})^{n-4}$, 
with $Q$ denoting the masses of the SM particles or the energy of the collider 
(see the above examples). This introduces a natural
hierarchy among the couplings that parametrize the NP effects at low energies.
In the second (non-linear) scenario the Higgs is very heavy or absent 
altogether. In that case the effective Lagrangian will be based on 
$D_\mu \Sigma$ instead of $D_\mu \Phi$. As such it will resemble the chiral
Lagrangian (\ref{chiral}), except for the fact that the weak and NP couplings 
are not neglected with respect to the scale that governs the 
strongly-interacting symmetry-breaking sector. In this way a hierarchy
emerges in powers of $Q/\Lambda_{NP}$ and/or $E^2/(4\pi v)^2$.
 
\subsection{Sensitivity to triple and quartic gauge-boson couplings at the NLC}

Up to now only the gauge-boson--fermion couplings and the (bi-linear) 
couplings between two gauge bosons are tested with high precision at 
low-energy experiments like LEP1 and the SLC. The presence of NP effects in 
these couplings is excluded below the per-cent level.
The natural next step would be to extend this to the non-abelian
triple and quartic gauge-boson couplings. Such high-precision tests 
should be seen in the light of the afore-mentioned effective Lagrangian for
NP effects, which will lead to specific contributions to the various couplings
between the gauge bosons. The contributions to the triple gauge-boson 
couplings are strictly the result of integrated-out NP loop effects (leading to
factors $1/16\pi^2$), whereas the contributions to the quartic gauge-boson 
couplings can also involve the exchange of NP particles. 

A completely general investigation, involving the simultaneous effects of all 
possible gauge-boson couplings, is not recommendable in view of the expected 
statistics at the NLC. For instance, from angular-momentum conservation one 
can infer the existence of 14 independent couplings of the type 
$WW\gamma$ and $WWZ$~\cite{Boudjema14}.\footnote{In the derivation of this 
number the scalar parts of the off-shell gauge bosons were neglected. This is 
motivated by the fact that the gauge bosons are in general coupled to 
approximately massless fermionic currents.} In the actual data analysis one 
is going to adopt a more pragmatic attitude by only considering those 
interactions that are most likely to show up in the data. This opens the way 
to a large variety of theoretically and experimentally motivated prejudices, 
reducing the number of independent couplings. There are three main guiding 
principles. First of all, the ordering of the anomalous operators according 
to their dimension can 
be exploited by only taking into account the operators with the lowest 
dimension. For the gauge-boson interactions this boils down to restricting the 
analysis to dimension-six operators. Secondly, the charge of the W is fixed 
and the photonic interactions are too well-established to tamper with. 
Therefore C or CP violating $WW\gamma$ interactions can be discarded, and  
$U(1)_{\mbox{\scriptsize em}}$ gauge invariance should be preserved. Thirdly, 
the low-energy experiments strongly constrain certain couplings. Consequently,
one should only consider operators that neither violate the custodial 
$SU(2)_V$ symmetry nor contribute to gauge-boson--fermion or
bi-linear gauge-boson interactions. Combining all this one 
ends up with only three independent triple gauge-boson couplings. For more 
information the reader is referred to the literature~\cite{TGCgeneral}.

\begin{figure}[tb]
  \begin{picture}(150,100)(0,-10)
    \ArrowLine(50,50)(25,25)        \Text(10,25)[lc]{$e^+$}
    \ArrowLine(25,75)(50,50)        \Text(10,75)[lc]{$e^-$}
    \Photon(50,50)(100,50){1}{9}    \Text(75,57)[bc]{$\gamma,Z$}
    \Photon(100,50)(125,25){1}{7}   \Text(104,28)[lc]{$W$}
    \Photon(100,50)(125,75){1}{7}   \Text(104,70)[lc]{$W$}
    \ArrowLine(150,10)(125,25)      \Text(165,10)[rc]{$\bar{f}_2'$}
    \ArrowLine(125,25)(150,40)      \Text(165,40)[rc]{$f_2$}
    \ArrowLine(150,60)(125,75)      \Text(165,60)[rc]{$\bar{f}_1'$}
    \ArrowLine(125,75)(150,90)      \Text(165,90)[rc]{$f_1$}
  \end{picture}
  \hfill
  \begin{picture}(150,100)(0,-10)
    \ArrowLine(25,80)(50,75)        \Text(10,80)[lc]{$e^-$}
    \ArrowLine(50,75)(50,25)        \Text(35,50)[lc]{$\nu_e$}
    \ArrowLine(50,25)(25,20)        \Text(10,20)[lc]{$e^+$}
    \Photon(50,75)(80,80){1}{7}     \Text(65,72.5)[tc]{$W$}
    \Photon(50,25)(80,20){1}{7}     \Text(65,27.5)[bc]{$W$}
    \ArrowLine(105,65)(80,80)       \Text(120,65)[rc]{$\bar{f}_1'$}
    \ArrowLine(80,80)(105,95)       \Text(120,95)[rc]{$f_1$}
    \ArrowLine(105,5)(80,20)        \Text(120,5)[rc]{$\bar{f}_2'$}
    \ArrowLine(80,20)(105,35)       \Text(120,35)[rc]{$f_2$}
  \end{picture}  
  \caption[]{The lowest-order `W-pair' diagrams contributing to 
             $e^+e^- \to f_1\bar{f}_1'f_2\bar{f}_2'$ in the SM. Here $f_i'$ 
             denotes the isospin partner of the fermion $f_i$. 
             On the left: the $s$-channel diagrams involving the triple 
             gauge-boson couplings. 
             On the right: the $t$-channel $\nu_e$-exchange diagram.}
  \label{eeWW}
\end{figure}
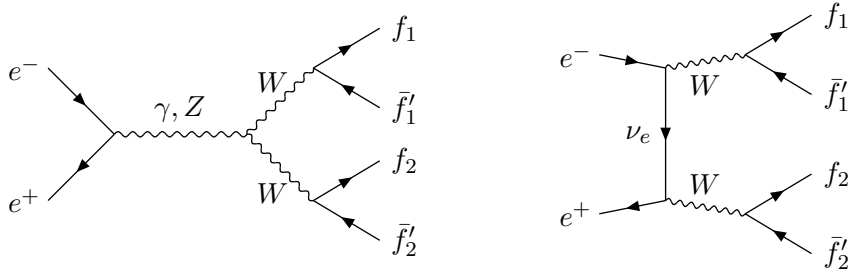

\begin{figure}[tb]
  \unitlength 1cm
  \begin{center}
  \begin{picture}(12,9.0)
  \put(0.7,6.5){\makebox[0pt][c]{$\sigma$}}
  \put(0.7,5.6){\makebox[0pt][c]{[pb]}}
  \put(9.0,0.3){\makebox{$\sqrt{s}$ \ [GeV]}}
  \put(9.3,3.4){$\sigma_{\mbox{\scriptsize born,SM}}$}
  \put(6.4,5.8){$\sigma_{\mbox{\scriptsize born},\,s}$}
  \put(5.3,6.5){$\sigma_{\mbox{\scriptsize born},\,t}$}
  \put(-0.8,-0.5){\includegraphics{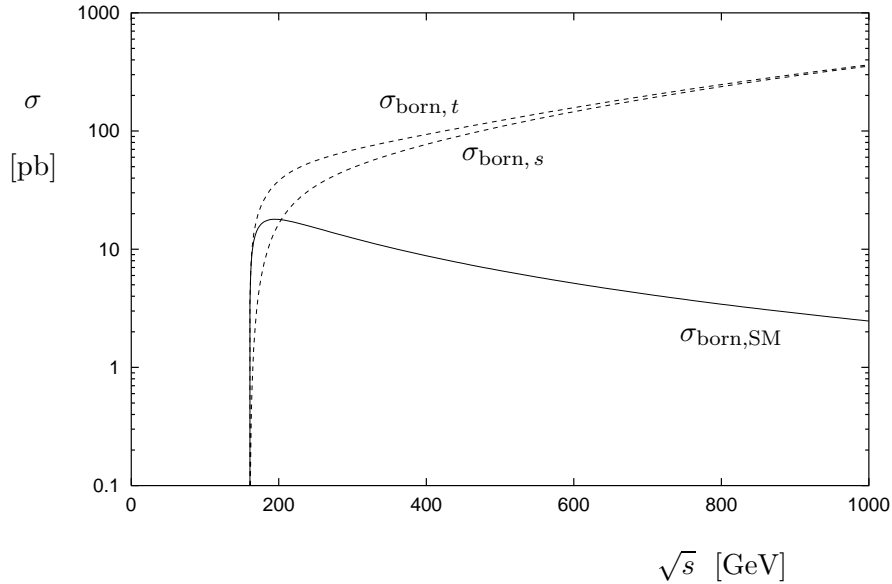}}
  \end{picture}
  \end{center}
  \caption[]{Gauge cancellations in the lowest-order SM process 
             $e^+e^- \to W^+W^-$. The dashed curves correspond to the Born 
             cross-sections arising from the $s$-channel 
             ($\sigma_{\mbox{\scriptsize born},s}$) and $t$-channel 
             ($\sigma_{\mbox{\scriptsize born},t}$) diagrams alone. 
             The solid curve corresponds to the complete Born
             cross-section ($\sigma_{\mbox{\scriptsize born,SM}}$).}
\label{WWcan}
\end{figure}

The main probe for anomalous triple gauge-boson couplings at the NLC will be 
the clean high-rate reaction $e^+e^- \to W^+W^-$ (see 
Fig.\,\ref{eeWW}).\footnote{In \cite{TGCBoudjema} it has been shown that the 
sensitivity only marginally degrades when going from the full process
$e^+e^-\!\to 4f$ to the `W-pair' process $e^+e^- \!\to $`$W^+W^-$'$\to 4f$, 
which can be obtained by imposing tight invariant-mass cuts on the decay 
products.}
Since the anomalous couplings are non-minimal, the delicate balance that is
present between the SM diagrams is broken. 
For longitudinal gauge bosons this upsets the gauge cancellations at high
energies (displayed in Fig.\,\ref{WWcan}), leading to deviations that are 
enhanced by factors of order $\sqrt{s}/M_{_W}$ for each longitudinal W boson.
Based on this observation and the scaling properties of the anomalous 
interactions involving transverse W bosons, the sensitivity to the anomalous
triple gauge-boson couplings is predicted to increase with the collider energy.
As a rule of thumb the sensitivity attainable at $e^+e^-$ colliders scales as 
$\sqrt{s L}$~\cite{TGCthumb}, where $L$ stands for the integrated luminosity. 
This automatically 
means that the NLC will do substantially better than LEP2 in view of the higher
energy and luminosity. The sensitivity to the s-channel diagrams involving the
triple gauge-boson couplings can be enhanced by cutting away the W bosons that
are produced in the forward direction, since the forward direction is 
completely dominated by the contributions from the $t$-channel
$\nu_e$ exchange. On top of that, in the clean environment of $e^+e^-$ 
collisions one can use the angular distributions of the decay 
products of the W bosons as polarization analyzers. In this way the 
sensitivity to specific polarization states and hence to specific
triple gauge-boson couplings can be enhanced. In order to fully exploit this 
opportunity it is important to identify the charge of at least one of the 
decay products of the decaying W bosons. At the NLC
the possibility of having polarized initial-state beams allows to
disentangle the effects from the $WW\gamma$ and $WWZ$ couplings. In this 
context also the other collider modes come in handy, making a whole host of 
other reactions accessible.  

Recent studies~\cite{TGCBoudjema,TGCBarklow} have shown that it will be 
feasible to probe the triple gauge-boson couplings at the per-mil level (at
500~GeV) or better (at 1.5~TeV). In this way the tests of the triple
gauge-boson interactions would be promoted to the level of high-precision 
measurements.
In order to test the quartic gauge-boson couplings one has to study 
gauge-boson fusion processes or processes that involve the production of three
gauge-bosons (like $e^+e^-\to W^+W^-\gamma,\,W^+W^-Z$). In view of the reduced 
amount of available phase space and the suppression by the additional powers 
of the electroweak couplings, it will be difficult to beat the LHC at this job.

\section{Accurate theoretical predictions in the SM}

In order to successfully achieve the physics goals at the NLC, a very accurate
knowledge of the SM predictions for the various observables is mandatory. 
This involves a proper understanding of radiative corrections as well as a 
proper treatment of finite-width effects. 

\subsection{The issue of gauge invariance}

As has become clear from the previous sections, the W-boson physics studies at
the NLC cover a large variety of processes with photons and/or fermions in the 
initial and final state. After all, the massive gauge bosons are unstable 
particles and can only be investigated through their decay products. 
If complete sets of graphs contributing to such a process are taken into
account, the associated matrix elements are in principle gauge-invariant.
However, the massive gauge bosons that appear as intermediate particles can 
give rise to poles $1/(k^2-M^2)$ if they are treated as stable particles. 
This can be cured by introducing the finite decay width in one way or another, 
while at the same time preserving gauge independence and, through a proper 
high-energy behavior, unitarity. In field theory, such widths arise naturally 
from the imaginary parts of higher-order diagrams describing the gauge-boson 
self-energies, resummed to all orders. This procedure has been used with great 
success in the past: indeed, the Z resonance can be described to very high 
numerical accuracy. However, in doing a Dyson summation of self-energy graphs, 
we are singling out only a very limited subset of all the possible higher-order
diagrams. It is therefore not surprising that one often ends up with a result 
that retains some gauge dependence. 

Till recently two approaches for dealing with unstable gauge bosons were 
popular in the construction of lowest-order Monte Carlo generators. The first 
one involves the systematic replacement $1/(k^2-M^2) \to 1/(k^2-M^2+iM\Gamma)$,
also for $k^2<0$. Here $\Gamma$ denotes the physical width of the gauge boson
with mass M and momentum $k$. This scheme is called the `fixed-width scheme'.
As in general the resonant diagrams are not gauge-invariant by themselves, 
this substitution will destroy gauge invariance. Moreover, it has no 
physical motivation, since in perturbation theory the propagator for 
space-like momenta does not develop an imaginary part. Consequently, 
unitarity is violated in this scheme. To improve on the latter another 
approach can be adopted, involving the use of a running width $iM\Gamma(k^2)$ 
instead of the constant one $iM\Gamma$ (`running-width scheme'). This,
however, still cannot cure the problem with gauge invariance. 

At this point one might ask oneself the 
legitimate question whether the gauge-breaking terms are numerically relevant 
or not. After all, the gauge breaking is caused by the finite decay width and
is, as such, in principle suppressed by powers of $\Gamma/M$. From LEP1 we 
know that gauge breaking can be negligible for all practical purposes. 
However, the presence of small scales can amplify the gauge-breaking terms.
This is for instance the case for almost collinear space-like photons or
longitudinal gauge bosons at high energies, involving scales of 
${\cal O}(p_{_B}^2/E_{_B}^2)$ (with $p_{_B}$ the 
momentum of the involved gauge boson). In these situations the external 
current coupled to the photon or to the longitudinal gauge boson becomes 
approximately proportional to $p_{_B}$. In other words, in these 
regimes sensible theoretical predictions are only possible if the amplitudes 
with external currents replaced by the corresponding gauge-boson momenta 
fulfill appropriate Ward identities.

In order to substantiate these statements, a truly gauge-invariant scheme is
needed. It should be stressed, however, that any such scheme is arbitrary to a 
greater or lesser extent: since the Dyson summation must necessarily be taken 
to all orders of perturbation theory, and we are not able to compute the 
complete set of {\it all} Feynman diagrams to {\it all} orders, the various 
schemes differ even if they lead to formally gauge-invariant results. Bearing 
this in mind, we need some physical motivation for choosing a particular 
scheme. In this context two options can be mentioned, which fulfill the 
criteria of gauge invariance and physical motivation. 

The first option is the so-called `pole scheme'~\cite{Veltman,Stuart,Aeppli}. 
In this scheme one decomposes the complete amplitude according to the pole 
structure by expanding around the poles, 
e.g.~$f(k^2)/(k^2-M^2) = f(M^2)/(k^2-M^2) + \mbox{finite~terms}$. As the 
physically observable residues of the poles are gauge-invariant, gauge
invariance is not broken if the finite width is taken into account in the pole 
terms $\propto 1/(k^2-M^2)$. It should be noted, however, that there exists 
some controversy in the literature~\cite{Aeppli,Stuart2} about the `correct' 
procedure for doing this and about the range of validity of the pole scheme, 
especially in the vicinity of thresholds. 

The second option is based on
the philosophy of trying to determine and include the minimal set of Feynman 
diagrams that is necessary for compensating the gauge violation caused by the 
self-energy graphs. This is obviously the theoretically most satisfying 
solution, but it may cause an increase in the complexity of the matrix 
elements and a consequent slowing down of the numerical calculations. For the 
gauge bosons we are guided by the observation that the lowest-order decay 
widths are exclusively given by the imaginary parts of the fermion loops in 
the one-loop self-energies. It is therefore natural to perform a Dyson 
summation of these fermionic one-loop self-energies and to include the other 
possible one-particle-irreducible fermionic one-loop corrections 
(`fermion-loop scheme')~\cite{BHF1}. For the process $e^+e^- \to 4f$ (see
Fig.\,\ref{eeWW}), this amounts to adding the fermionic triple gauge-boson 
vertex corrections. The complete set of fermionic contributions forms a 
gauge-independent subset and obeys all Ward identities exactly, even with 
resummed propagators~\cite{BHF2}. 
As mentioned above, the validity of the Ward identities guarantees a proper 
behavior of the cross-sections in the presence of collinear photons and at 
high energies in the presence of longitudinal gauge-boson modes. On top of 
that, within the fermion-loop scheme the appropriately renormalized matrix 
elements for the generic process $e^+e^- \to 4f$ can be formulated in 
terms of effective Born matrix elements, using the familiar language of 
running couplings~\cite{BHF2}.

A numerical comparison of the various schemes~\cite{BHF1,BHF2} confirms the 
importance of not violating the Ward identities. For the process 
$e^+e^- \to e^-\bar{\nu}_e\,u\bar{d}$, a process that is particularly 
important for studying triple gauge-boson couplings, the impact of violating 
the electromagnetic $U(1)_{\mbox{\scriptsize em}}$ gauge invariance was 
demonstrated in~\cite{BHF1}. Of the above-mentioned schemes only the 
running-width scheme violates $U(1)_{\mbox{\scriptsize em}}$ gauge invariance. 
The associated gauge-breaking terms are enhanced in a disastrous way by a 
factor of ${\cal O}(s/m_e^2)$, in view of the fact that the electron may emit 
a virtual (space-like) photon with $p_\gamma^2$ as small as $m_e^2$. A similar 
observation can be made at high energies when some of the 
intermediate gauge bosons become effectively longitudinal. There too the 
running-width scheme renders completely unreliable results~\cite{BHF2}. In 
processes involving more intermediate gauge bosons, e.g.~$e^+e^- \to 6f$
(see Fig.\,\ref{LLtoLL}a),
also the fixed-width scheme breaks down at high energies as a result of
breaking $SU(2)_L$ gauge invariance.

\subsection{Radiative corrections in the SM}

By employing the fermion-loop scheme all one-particle-irreducible fer\-mionic 
one-loop corrections can be embedded in the tree-level matrix elements. This 
results in running couplings, propagator functions, vertex functions, etc. 
However, there is still the question about the bosonic corrections. Such 
corrections might mimic the presence of anomalous gauge-boson interactions if
they are not taken into account properly.\footnote{Note that the SM 
predictions need not have a $0.1\%$ precision in order to study 
anomalous triple gauge-boson couplings at the $10^{-3}$ level. These anomalous
couplings are non-minimal and therefore lead to enhanced effects.} A large 
part of the bosonic corrections, as e.g.~the leading QED corrections, 
factorize and can be treated by means of a convolution, using the 
fermion-loop-improved cross-sections in the integration kernels (see 
e.g.~appendix A of \cite{LEP2WW}). This allows 
the inclusion of higher-order QED corrections and soft-photon 
exponentiation. In this way various important effects can be covered. 
In this respect especially the emission of hard photons from the initial state
is noteworthy~\cite{ISRboost,WWreview}. The associated hard-photon boost 
effects will lead to a
redistribution of phase space, which affects the angular distributions.
Also the polarization of the produced gauge bosons is affected by the presence
of such boosts. The best-suited observables for probing the NP sector normally
involve small cross-sections and are therefore extremely sensitive to 
redistribution effects. For instance, observables related to longitudinal gauge
bosons will receive large corrections from the tranverse modes. 
In order to reduce the effects from the hard photons to a minimum, appropriate
cuts on the total final-state energy and momentum have to be imposed.

Only taking into account the (leading) factorizing bosonic corrections is not
sufficient. The remaining bosonic corrections can be large, especially at high 
energies where logarithmic corrections $\propto \log^2(M_{_{W,Z}}^2/s)$ 
emerge~\cite{WWreview,largelogs}. In order to include the remaining bosonic  
corrections one might attempt to extend the 
fermion-loop scheme. In the context of the background-field method a Dyson 
summation of bosonic self-energies can be performed without violating
the Ward identities~\cite{BFM}. However, the resulting matrix elements depend
on the quantum gauge parameter at the loop level that is not completely 
taken into account. As mentioned before, the perturbation series has to be 
truncated; in that sense the dependence on the quantum gauge parameter could be
viewed as a parametrization of the associated ambiguity. 

As a more appealing strategy one might adopt a hybrid scheme, adding the 
remaining bosonic loop corrections by means of the pole scheme. This is 
gauge-invariant and contains the well-known bosonic corrections for the 
production of on-shell gauge bosons (in particular W-boson 
pairs~\cite{WWRC1,WWRC2}). Moreover, 
if the quality of the pole scheme were to degrade in certain regions of 
phase-space, the associated error is reduced by factors of $\alpha/\pi$. 
It should be noted that the application of the pole scheme to photonic 
corrections requires some special care, because in that case terms 
proportional to $\log(k^2-M^2)/(k^2-M^2)$ complicate the pole
expansion~\cite{Aeppli,WWreview}.

\begin{center}
  {\large\bf Acknowledgements}
\end{center}

I would like to thank W.L.~van~Neerven for various clarifying discussions.
Also the financial support by the TEMPUS program is gratefully acknowledged.
Special thanks go to the organization of the Cracow Epiphany Conference, to
the members of the physics institute of the Jagellonian University, and to
the members of the INP for their warm hospitality.

\end{document}